# Stabilization of multiple emulsions using natural surfactants

H. Ghasemi [1], H. Mazloomi [2], and H. Hajipour[2,*]

[1]School of Materials Engineering, College of Engineering, University of Tehran, P.O. Box 11365 4563, Tehran, Iran

[2]School of Chemical Engineering, College of Engineering, University of Tehran, P.O. Box 11365 4563, Tehran, Iran

**Abstract**

In an emulsion system, emulsifier is one of the most important substances as it determines the formation, stability and physicochemical properties of emulsions.
In this study, the effects of emulsifier concentration, type of hydrophilic emulsifier, as well as portions of primary emulsion (weight) on the stability of W/O/W emulsions were investigated. Microscopy images of W/O/W emulsions indicated that the emulsions prepared with 0.5 gram of sodium caseinate have superior stability over other synthesis conditions. Finally, emulsions were prepared using different types of emulsifier (NaCN, Cremophor, Tween 60). Our results showed that emulsions made form Cremophor and Tween 60 in comparison with sodium caseinate possess smaller droplets size with enhanced stability.

## 1- Introduction

An emulsion is a colloidal dispersion of two immiscible liquids, such as oil and water, in which one liquid is dispersed in another [1]. There are two types of simple emulsion systems, oil-in-water (O/W) emulsions and water-in-oil (W/O) emulsions. In O/W emulsions, oil is dispersed in a continuous water phase, whereas in W/O emulsion, water is the dispersed phase in an oil phase [2].
Multiple emulsions are more complex systems, in which the dispersed phase is itself an emulsion; these emulsions can be classified into two major types: water-oil-water (W/O/W) emulsions and oil-water-oil (O/W/O) emulsions [2].
A W/OW emulsion is an O/W emulsion in which the dispersed oil phase is a water-in-oil (W/O) emulsion [3].
There are two main methods for the formation of W/O/W emulsion: one step emulsification and two step emulsification. One step emulsification methods include strong mechanical agitation and phase inversion. A W/O emulsion is formed initially, but part of the emulsion inverts and forms a W/O/W emulsion. Two–step emulsification methods involve making a fine primary W/O emulsion and then dispersing the primary emulsion in a solution with a hydrophilic emulsifier [4]. A critical factor that determines the effectiveness of this method is intrinsic stability of the internal W/O emulsion. It is important that the presence of hydrodynamic perturbations during the second emulsification stage does not lead to any significant breakdown of the primary emulsion [5].

Various techniques have been applied in two-step methods, including blender and mixer, high-pressure homogenizer, microfluidiser, membranes ets. The different techniques not only cause the reduction in droplet size but also influence the adsorption of the emulsifier [6, 7]. High- pressure homogenisers can produce very fine droplets, which cannot be produce by mixers or rotor-stator homogenisers [8]. The difference in droplet size is directly correlated to the energy density. The higher the energy density, the smaller are the droplets. The energy input is about $10^{12}$ w/m$^3$ for the high – pressure homogenizer [8].

After formation of emulsion, the stability of the disrupted droplets is determined by the correct selection of the composition of W/O/W emulsions. The main constituents of a W/O/W emulsion are the internal aqueous phase, hydrophobic emulsifiers, the oil phase, hydrophilic emulsifiers and the external aqueous phase [9, 10].

The aqueous phase is the dispersed phase in a W/O emulsion and the continuous phase in a W/O/W emulsion. Internal aqueous phases are often solutions of encapsulated compounds, such as sugar, salt and nutrients. External aqueous phases are solutions of emulsifiers (e.g. proteins) and stabilizers [11].

A stable W/O emulsion is essential to the stability of the secondary W/O/W emulsion, and it is strongly correlated to the hydrophobicity of the oil phase. High hydrophobicity materials such as mineral oils or hydrocarbon solvents are commonly used as the oil phase in studies of W/O or W/O/W emulsion [12].

Emulsifiers are amphiphilic compounds that possess two distinct groups in the same molecule a hydrophobic group, which has an affinity for the oil phase, and a hydrophilic group, which has an affinity for water. Emulsifiers lower the interfacial tension and facilitate droplet disruption, resulting in smaller droplets. An emulsifier determines which phase is the continuous phase and which is the dispersed phase; use the wrong emulsifier could result in an inverted emulsion [12, 13].

PGPR is derived from castor oil is known to be one of the most efficient oligomeric emulsifiers for W/O emulsions and is commonly used polymeric hydrophobic emulsifiers in W/O/W emulsions [12].

The ability of sodium caseinate (NaCN), a milk protein, to act as an emulsifier and stabilizer at oil-water interfaces is well documented. NaCN is amphiphilic proteins with a strong tendency to adsorb at oil-water interfaces during emulsion formation, reducing interfacial tension. This produces an adsorbed layer of protein around the oil droplets, which protects them against subsequent coalescence and flocculation [14].

Jahaniaval et al. were studied the physicochemical (solubility and hydrophobicity), and functional (emulsifying activity index and emulsifying capacity) properties of soluble sodium caseinate fractions as a function of pH (3–8) and temperature (50–100°C). They showed that the soluble protein fractions from sodium caseinate heat treated near the PI of the caseins, to have enhanced emulsifying activity and capacity [15].

Dickinson et al. were investigated the effect of the combination of ionic calcium and non-ionic surfactant (Tween 20) on the visual creaming behavior and rheology of n-tetradecane-in-water emulsions (4 wt% protein, 30 vol% oil, mean droplet diameter 0.4 μm) prepared at pH 6.8 with sodium caseinate. Their results showed that at low R (calcium/caseinate molar Ratio), displacement from the interface of calcium-bound protein provides an extra source of (mainly protein-bound) ionic calcium in the aqueous phase, leading to fewer and larger caseinate aggregates that are then incapable of inducing depletion flocculation. At high R, displacement of protein from the oil–water interface results in emulsion restabilization through the disruption of calcium-induced caseinate bridges between the droplets [16]. Furthermore, it has been shown in several studies that fine residues can affect the rheology of complex fluids and as a result the stability of W/O [17, 18]. For an instance, Mozaffari et al. using a novel nanofluidic platform showed that the asphaltenic aggregates can alter the oil rheology from Newtonian to non-Newtonian with small yield stress [17]. The presence of small yield stress can prevent the rupture of oil film and therefore enhance the stability of W/O.

In this study, the effect of different concentrations of Sodium Caseinate (0.5, 1, 1.5 gr) in the external aqueous, different portions of primary emulsion (25%, 40%, 50%) and type of hydrophilic emulsifiers on the formation and stability of W/O/W emulsions prepared using a Vibracell Ultrasound was investigated.

## 2- Materials and Methods

### 2-1- Emulsion preparation

W/O/W emulsions were prepared by a two-step process. Water and sodium chloride (41mg) were dispersed in soybean oil (60 gr) containing PGPR (1.5 gr) and HMPC (200 mg), and the mixtures were homogenized at 700W using a Vibracell Ultrasound at $37^{0C}$ for 30 minutes to obtain the primary W/O emulsions. The second step involved the dispersion of the primary emulsions into solutions containing hydrophilic emulsifiers (NaCN, Tween 60, Cremophor), using the same homogenizer.

### 2-2- Confocal laser microscopy

Confocal scanning laser microscopy (CSLM) was used to observe the multiple emulsion samples.

### 2-3- Determination of average droplet size of W/O/W emulsions

The droplet sizes of W/O/W emulsions were measured using Malven MasterSizer MSE.

## 3- Results and discussion

### 3-1- Effect of NaCN concentration in the external aqueous on the Formation W/O/W emulsions

NaCN is not monomeric but is aggregated to some extent in solution. The nature and the size of the aggregates are probably dependent on protein concentration, temperature, presence of ions and processing history. During the dynamic conditions of homogenization, much of the protein material is transported to the oil-water interface by convection rather than by diffusion. The rate of adsorption of protein is determined by the size of the adsorbing proteins or aggregates and the immediately available binding sites on the molecule. NaCN is used at lower concentration, compared with monomeric surfactants, to avoid depletion flocculation of emulsions containing unabsorbed NaCN.

Figure 1 shows Confocal microscopy images of W/O/W emulsions prepared with different concentration (0.5, 1, 1.5 gr) of NaCN in the external aqueous using Vibracell Ultrasound homogenizer.
Comparing three concentration of NaCN in the external aqueous, samples with o.5 gr of NaCN were found to be the most stable emulsions.

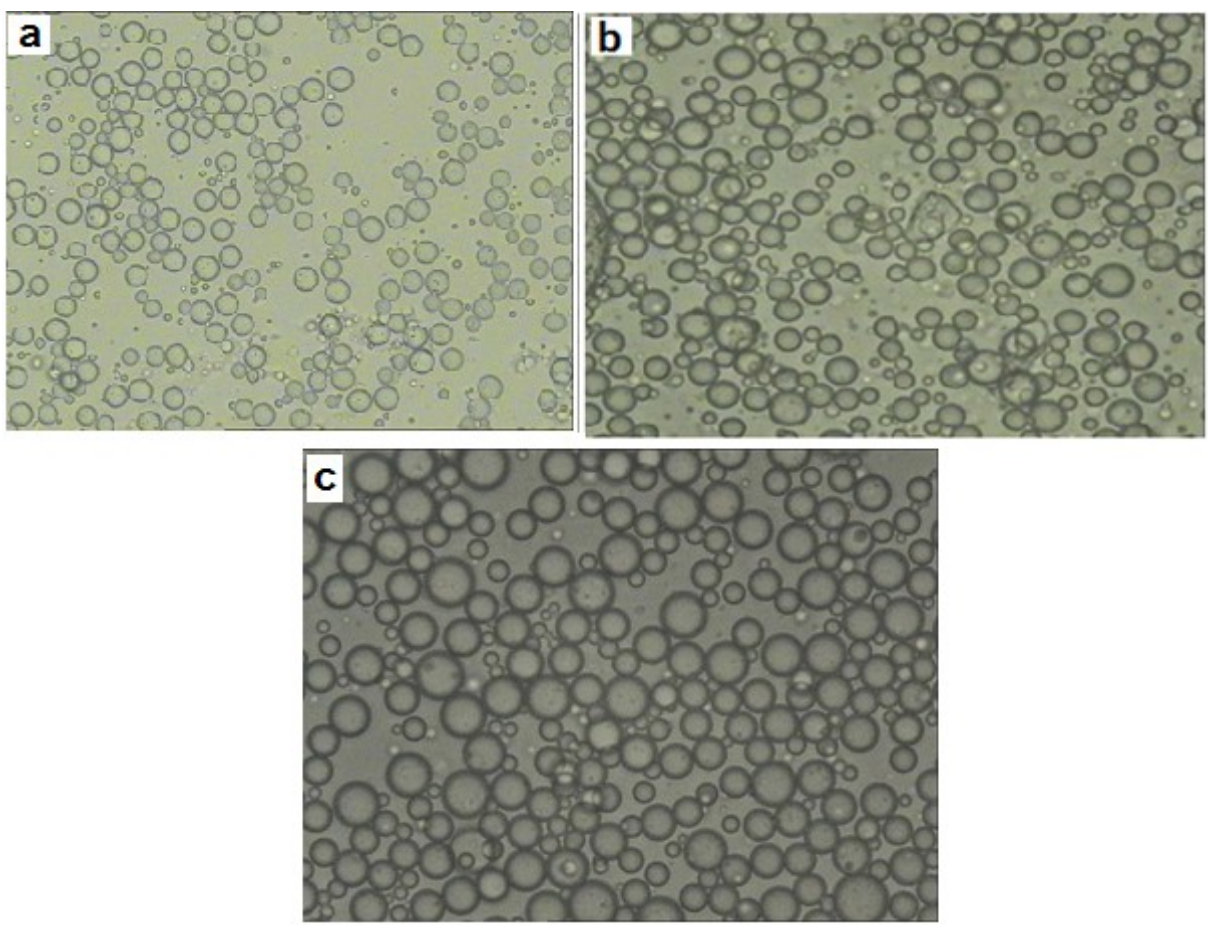

**Fig1**: Confocal microscopy images of prepared W/O/W emulsions with different concentration of NaCN in the external aqueous a) 0.5 g NaCN  b) 1g  NaCN  c) 1.5g  NaCN

### 3-2-   Effect of portions of primary emulsion on the Formation of W/O/W emulsions

The increase of water phase induced greater instability, possibly due to increased incidence of coalescence and bridging at higher water volume, both of which led to reduction in total water droplet surface area.
Figure 2 shows Confocal microscopy images of W/O/W emulsions prepared with different portions of primary emulsion (25%, 40%, 50%) using Vibracell Ultrasound homogenizer.

Comparing three portions of primary emulsion, samples with 50% W/O and 50 % secondary water phase found to form small droplets with superior stability over long period of time.

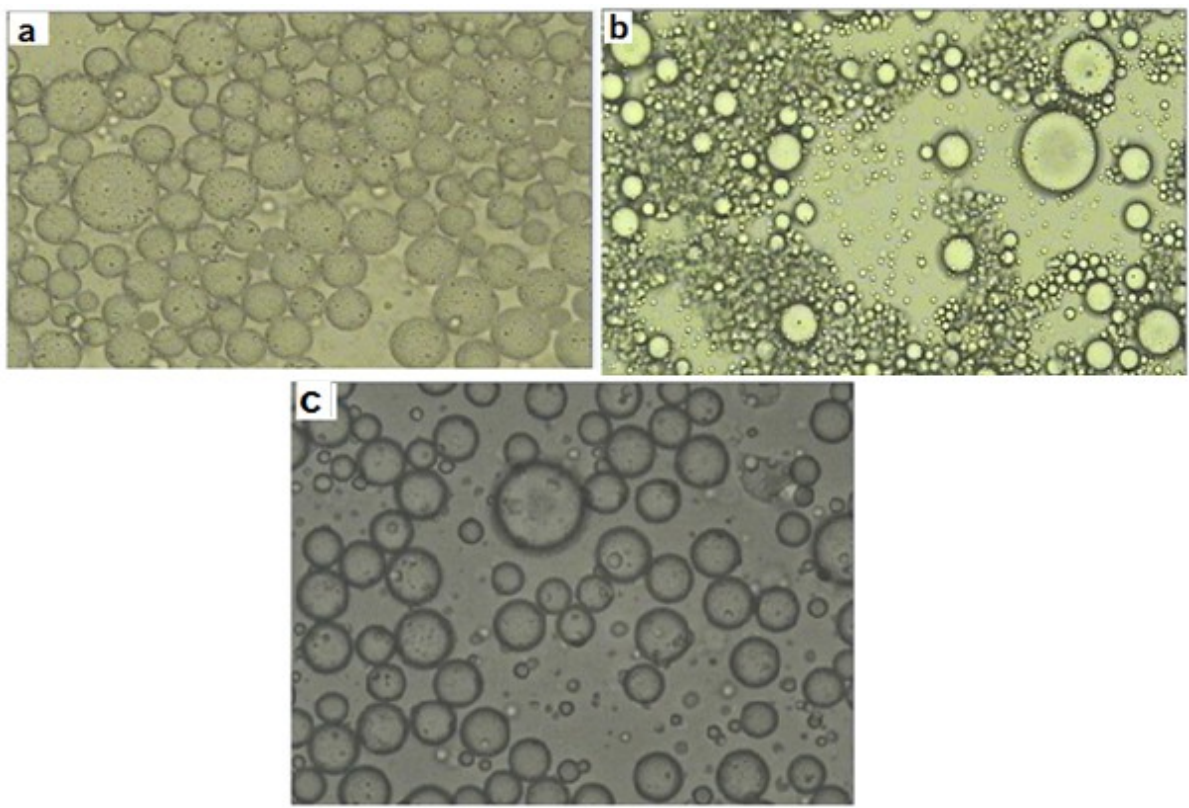

**Fig2**: Confocal microscopy images of prepared W/O/W emulsions with different portions of primary emulsion, a) 50% W/O, b) 40%, W/O, c) 25% W/O

### 3-3 - Effect of hydrophilic emulsifiers on droplet sizes of W/O/W emulsions

Hydrophilic emulsifiers contributed mainly on the stability of oil droplets, an appropriate amount of hydrophilic emulsifiers could prevent oil droplets from coalescence, therefore the droplet size distribution would not undergo significant change for a long period of time. Insufficient hydrophilic emulsifiers lead to larger oil droplets in emulsion as a result of coalescence.

Proteinaceous emulsifiers such as NaCN are flexible in structure. When used to form emulsions, NaCN adsorbs to the oil-water interface in an unfolded way to gives a high surface coverage and large adsorbed layer dimension [19]. Tween 60 and Cremophor are natural emulsifiers which are promising materials for food and medical applications.

Figure 3 shows the comparison droplet size distribution of W/O/W emulsions prepared with different hydrophilic emulsifiers (Cremophor, Tween 60, NaCN) using Vibracell Ultrasound homogenizer.

Comparing three hydrophilic emulsifiers, samples with Cremophor & Tween 60 had a narrow droplet size distribution and a smaller mean droplet size.

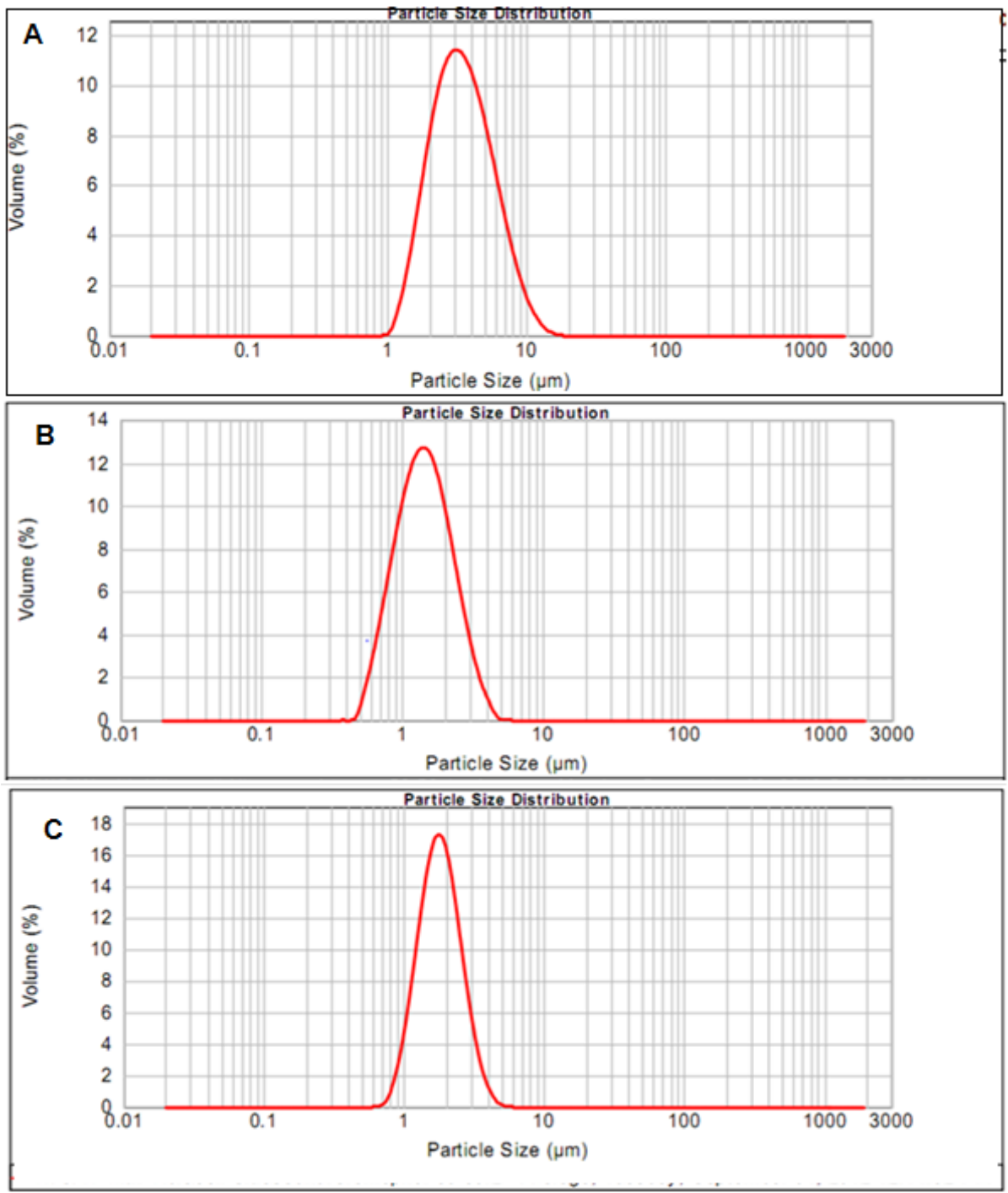

**Fig 3**: droplet size distribution of W/O/W emulsion with different hydrophilic emulsifier. The external phase contained a) 0.5 gr NaCN, b) 0.5 g tween 60, c) 0.5 g Cremophor. Emulsions were prepared with 50% portions of primary emulsion using Vibracell Ultrasound homogenizer.

## 4- Conclusions

In this study, the effects of emulsifier concentration, type of hydrophilic emulsifier, as well as portions of primary emulsion (weight) on the stability of W/O/W emulsions were investigated. By comparing the results obtained from different portions of primary emulsions (50%, 40% and 25% (w/w)), it can be inferred that multiple emulsion prepared by 50% (w/w) W/O contains more water droplets with enhanced stability. Emulsions were prepared using different types of emulsifier (NaCN, Cremophor, Tween 60). It was found that emulsions made by Tween 60 and Cremophor form smaller droplets which in contrast with those obtained from NaCN. Tween 60 and Cremophor are natural emulsifiers which can be used for food and medical applications.

**Acknowledgements**

The authors are grateful to the Materials Research Institute (MRI) for providing us with the financial support. We also express our appreciation to Dr. Pishvai for his invaluable feedback.

## References

1. McClements DJ. 2005. Food Emulsions: Principles, Practices and Techniques. Boca Raton, FL:CRC Press.
2. Pradhan, M., & Rousseau, D. (2012). A one-step process for oil-in-water-in-oil double emulsion formation using a single surfactant. Journal of Colloid and Interface Science 386, 398–404.
3. Sagalowicz, L., & Leser, M. E. (2010). Delivery systems for liquid food products. Journal of Colloid & Interface Science, 15(1e2), 61e72.
4. Benichou, A., Aserin, A. and Garti, N. (2001). Polyols, high pressure, and refractive indices equalization for improved stability of W/O emulsions for food applications. Journal of Dispersion Science and Technology 22(2&3): 269-280.
5. Van der Graaf, S., Schroen, C. and Boom, R. (2005). Preparation of double emulsions by membrane emulsification, a review. Journal of Membrane Science 251: 7-15.
6. Liu, E. and McGrath, K. (2005). Emulsion microstructure and energy input, roles in emulsion stability. Colloids and surfaces A: Physicochemical and Engineering Aspects 262:101-112.
7. Perrier-Cornet, J. M., Marie, P. and Gervais, P. (2005). Comparision of emulsification efficiency of protein-stabilized oil- in- water emulsions using jet, high pressure and colloid mill homogenization. Journal of Food engineering 66 (2): 211-217.
8. Schubert, H. and Engel, R. (2004). Product and formulation engineering of emulsions. Trans IchmE, PartA: Chemical Engineering Research and Design 82 (A9): 1137-1143.
9. Taisne, L., Walstra, P. and Cabane, B. (1996). Transfer of oil between emulsion droplets. Journal of Colloid and interface Science 184: 378 – 390.

10. Keshmiri, K., Mozaffari, S., Tchoukov, P., Huang, H. and Nazemifard, N. (2016). Using Microfluidic Device to Study Rheological Properties of Heavy Oil. arXiv preprint arXiv:1611.10163.
11. Garti, N. and Aserin, A. (1996b). Pharmaceutical emulsions, double emulsions and microemulsions. In Microencapsulation–Methods and Industrial Applications. Ed. Benita, S. New York, Marcel Dekker: 411-534.
12. Benichou,A., Aserin, A. and Garti, N. (2001). Polyols, high pressure, and refractive indices qualization for improved stability of W/O emulsions for food applications. Journal of Dispersion Science and Technology 22 (2&3): 269-280.
13. Kanouni, M., Rosano, H. and Naouli, N.(2002). Preparation of stabl double emulsion (W1/O/W2): role of the interfacial films on the stability of the system. Advances in Colloid and interface Science 99: 229-254.
14. Ye, A. and Singh, H. (2001). Interfacial composition and stability of sodium caseinate emulsions as influenced by calcium ions. Food Hydrocolloids 15 (2): 195-207.
15. Jahaniaval F., Kakuda Y., Abraham V. and Marcote M.F., Soluble protein fraction from pH and heat treated sodium caseinate: physicochemical and functional properties. Food Research International. 33: 637-647, 2008.
16. Dickinson, E., Radford, S.J. and Golding, M., Stability and rheology of emulsions containing sodium caseinate: combined effects of ionic calcium and non-ionic surfactant. Food Hydrocolloids. 16: 153-160, 2003.
17. Mozaffari, S., Tchoukov, P., Mozaffari, A., Atias, J., Czarnecki, J. and Nazemifard, N. (2017). Capillary driven flow in nanochannels–Application to heavy oil rheology studies. Colloids and Surfaces A: Physicochemical and Engineering Aspects, 513, pp.178-187.
18. Mozaffari, S., Tchoukov, P., Atias, J., Czarnecki, J. and Nazemifard, N. (2015). Effect of asphaltene aggregation on rheological properties of diluted athabasca bitumen. Energy & Fuels, 29(9), pp.5595-5599.
19. Euston, S. and Hirst, R. (2000). The emulsifying properties of commercial milk protein products in simple oil-in-water emulsions and in a model food system. Journal of Food Science 65 (6): 2000.